\documentstyle[aasms4]{article}

\newcommand{\kms}{$\rm km \ s^{-1}$}

\newcommand{\amin}{$^{\prime}$}                     
\newcommand{\asec} {$^{\prime\prime}$}                     
\newcommand{\adeg} {$^{\circ}$}                      
\newcommand{\rasec}{\hbox{$\,$\raise 0.6 ex \hbox{\rm s}\kern-.35em
                  \lower 0.0 ex \hbox{.}$\,$}}        
\newcommand{\decsec}{\hbox{$\,$\raise 0.0 ex \hbox{$\asec$}\kern-.45em
                  \lower 0.0 ex \hbox{.}$\,$}}         
\newcommand{\decmin}{\hbox{$\,$\raise 0.0 ex \hbox{$\amin$}\kern-.45em
                  \lower 0.0 ex \hbox{.}$\,$}}         
     
\newcommand{\gtabouteq}{\,\hbox{\raise 0.5 ex \hbox{$>$}\kern-.77em 
                    \lower 0.5 ex \hbox{$\sim$}$\,$}}       
\newcommand{\ltabouteq}{\,\hbox{\raise 0.5 ex \hbox{$<$}\kern-.77em 
                     \lower 0.5 ex \hbox{$\sim$}$\,$}}

\newcommand{\Mo} {~M_{\odot}}
 
\begin{document}
 
\slugcomment{Submitted to Ap.~J.~L.}
 
\lefthead{English, Taylor, Mashchenko, Irwin, Basu and Johnstone}
\righthead{}
 
\title{Galactic Worm 123.4-1.5: 
A Mushroom-shaped HI Cloud}

\author{Jayanne English\altaffilmark{1,2}, 
A.R. Taylor\altaffilmark{3},
S.Y. Mashchenko\altaffilmark{4}, Judith A. Irwin\altaffilmark{1}, 
Shantanu Basu\altaffilmark{5}, Doug Johnstone\altaffilmark{6}}

\altaffiltext{1}{Queen's University, Dept. of Physics, Kingston, 
Ontario, Canada, K7L 3N6}
\altaffiltext{2}{Space Telescope Science Institute, Baltimore, MD}
\altaffiltext{3}{University of Calgary, Dept. of Physics and 
Astronomy, Calgary, Alberta, Canada T2N 1N4}

\altaffiltext{4}{Universit\'e de Montr\'eal, Dept. de Physique, 
Montr\'eal, Qu\'ebec, 
Canada H3C 3J7}

\altaffiltext{5}{University of Western Ontario, 
Dept. of Physics and Astronomy, London, Ontario, 
Canada N6A 3K7}

\altaffiltext{6}{University of Toronto, 
 60 St. George Street, Toronto, Ontario,
 Canada M5S 3H8}

 
\begin{abstract}
  
  The Dominion Radio Astrophysical Observatory's Synthesis
  Telescope provides the highest resolution data (1 arcmin and
  0.82 \kms) to date of an H~I worm candidate. Observed as part
  of the Canadian Galactic Plane Survey, mushroom-shaped GW
  123.4-1.5 extends only a few hundred parsecs, contains
  $\sim10^5 \Mo$ of neutral hydrogen, and appears unrelated to a
  conventional shell or chimney structure.  Our preliminary
  Zeus-2d models use a single off-plane explosion with a modest
  ($\sim 10^{51}$ ergs) energy input. These generic simulations
  generate, interior to an expanding outer blast wave, a buoyant
  cloud whose structure resembles the morphology of the observed
  feature.  Unlike typical model superbubbles, the stem can be
  narrow because its width is not governed by the pressure behind
  the blast wave nor the disk scale height. Using this type of
  approach it should be possible to more accurately model the
  thin stem and other details of GW 123.4-1.5 in the future.
 
Subject Headings:  ISM: bubbles --- ISM: individual (GW 123.4-1.5)
--- ISM: structure --- Galaxy: halo --- Galaxy: structure

\end{abstract}
 
\section{Introduction}
\label{intro}

\nocite{heil84}Heiles (1984) has identified atomic
hydrogen (HI) gas filaments ``crawling'' away from the plane of the
inner Galaxy.  These so-called ``worms'' were proposed to be parts of
larger HI shells blown by the energetic stellar winds or supernovae in
stellar associations.  Such open, or blown out, shells would serve as
conduits for hot gas and radiation to escape
into the galactic halo, as recently
observed in the superbubble/chimney reported by
\nocite{NorTayDew96} Normandeau et~al.~1996.  Koo et al. (1992) later 
produced a catalogue of 118 Galactic worm candidates,
defining as a worm any dusty, H~I structure perpendicular to the
Galactic Plane.

The Canadian Galactic Plane Survey (CGPS) is currently
mapping a 70 degree longitude segment of the northern Galaxy
at high resolution in HI (\nocite{taylor99} Taylor et al. 1999).
One Galactic worm candidate, GW 123.4-1.5, is
within the early regions surveyed.  These
observations reveal that GW 123-1.5 is an unusual,
mushroom-shaped cloud, hundreds of pc in size,
apparently unrelated to a conventional
shell or chimney structure.
In this paper we present the observations,
derive the observed properties of the mushroom cloud,
and discuss possible scenarios for its origin.

\section{Observations} 
\label{obs}

The HI emission from GW 123.4-1.5 was observed using the Synthesis
Telescope of the Dominion Radio Astrophysical Observatory.  A complete
synthesis of the field of view (2.5$^{\circ}$ at $\lambda$21 cm) was
combined with short spacing information from the DRAO 26-m telescope
(\nocite{Higgs99}Higgs 1999) so that there are no gaps in the uv plane
and all spatial scales are detected to a limiting resolution of $1'
\times 1.14'$ (RA $\times$ DEC).  Data cubes obtained for each of
10 fields were mosaiced together.  Five northern fields within the
mosaic, observed between 1995 and 1997, are part of the on-going
CGPS and five southern fields were observed in June 1997 and July
1998 with identical parameters by extending the CGPS grid to more
negative Galactic latitude.

The complex gain calibration was applied from observations of
strong unresolved sources before and after each 12-hour
synthesis. The S7 region, with an adopted brightness temperature
of 100 K, was used as the flux calibrator for the 26-m
observations.  Each cube contained 256 velocity channels with a
channel spacing of 0.824 km~s$^{-1}$ and an LSR velocity range of
$-164.7$ to 58 km~s$^{-1}$.  Each field was then corrected for
primary beam attenuation, resulting in a non-uniform noise
distribution in the mosaiced images. The mosaic's r.m.s. noise
per channel ranges from a minimum of 2.9~K (at the field center
locations) to 4.6~K.

A map of HI emission integrated over the LSR velocity range
$-31.1$ \kms\, to $-43.5$ \kms, is shown in Figure 1a.  The image
shows the mushroom cloud extending vertically out of the galactic
plane toward negative Galactic latitudes.  The ``stem'' of the
mushroom extends about 3\adeg\, out of the plane into a mushroom
``cap'' that is about 2.5\adeg\, wide. At the bottom outside
edges of the cap, two ``lobes'' extend back toward the plane on
either side of the stem.  At the top of the cap, wispy filaments
reach toward larger latitudes.  We have not assumed apparently
independent H~I features are part of the cloud.  For example,
the ring at $(l,b)=(123.2,-6.2)$ is atomic gas surrounding
the HII region S184, which is at a distance of 2.2 $\pm$ 0.7 kpc
(Fich, private communication), and it would be speculative to assume
this region is a component of the mushroom cloud.

\section{Derived Properties of the Mushroom Cloud}
\label{mush}

The velocity-latitude diagram in Figure 1b, taken along
longitude, -123.6\adeg\, which cuts through the stem and central
cap (see slice in Figure 1a), shows that the stem emerges from
the ambient HI in the mid-plane at a velocity of about $-43$
\kms.  Using the galactic rotation model of \nocite{brbl93}Brand
\& Blitz (1993 and presented in \nocite{BBur88}Burton 1988), the
kinematic distance is 3.8 $\pm$ 1.2 kpc, placing the cloud in the
inner edge of the Perseus arm.  At this distance, the mushroom
cloud extends from $|z| \approx 70$ pc (where it becomes distinct
from the disk emission) to 420 pc, for a total projected length
perpendicular to the plane of 350 pc, about three times larger
than the estimate of \nocite{khr92}Koo~et~al.(1992).

The velocities along the stem become less negative as $|z|$ increases
(Figure 1b; clearer when stepping through the cube).  The stem
overlaps with the cap starting at about $|z| \sim 200$ pc and $v_{\rm
  LSR} = -38$ \kms, then extends an additional $\sim$80~pc into the
cap.  The stem is typically 35-40 pc wide and appears to have a cavity
as well as a velocity signature indicating either expansion,
contraction or helical motion.

The central portion of the cap has little or no velocity gradient with
respect to $|z|$ (Figure 1b)
or longitude (Figure 1d).  However, Figure 1c shows that the two
lobes of the cap that extend back toward the plane are blueshifted
with respect to the central cap region by 5 \kms.  In contrast, the
wispy diffuse southeast parts of the cap are redshifted with respect
to the main body of the cap.

The velocity characteristics, presented in
Table~\ref{tablepar}, were measured for the stem component, the cap
subcomponents (lobes, central region and wisps), and the entire
mushroom cloud.  Velocity characteristics were obtained from
brightness temperature ($\rm T_B$) versus velocity profiles averaged
over visually selected regions of each component.  The errors reflect
the values measured for different baseline estimates and different
box sizes surrounding the emission regions.

The mass estimates in Table 1 were obtained from 
column densities, integrated over
the total velocity range for that component and
averaged over the component spatially.  These were also corrected
for an average ``background'' column density.
The mean excess column density of the mushroom
above the background, over the velocity range of the 
entire structure, is $N_H = 6 \times 10^{20}$ cm$^{-2}$, roughly
equal to the average background over the same velocity range.  The
total hydrogen mass of the cap is then $\sim10^5 \Mo$, and the average
density of hydrogen atoms within the cap is $\sim0.2$ cm$^{-3}$.  The
density in the ambient medium surrounding the mushroom is less well
determined, since the distance of the column over the same velocity
interval is not known.  However the column is not likely to exceed
about 1 kpc (compared to about 200 pc for the mushroom), so the
over-density between the mushroom and surrounding gas is not likely to
exceed a factor of 5.

Using half the visually determined velocity range as an estimate
of the average internal motion of each component, the internal
kinetic energy of the hydrogen gas in the cap is $1 \times
10^{50}$ ergs (note: the velocity dispersion determined from
the $\rm T_B$ profile gives an energy estimate of a couple $\times
10^{49}$ ergs).  The radial velocity difference between the base
of the stem and the mid-region of the cap is about 7 \kms\, and
the bulk kinetic energy of the cloud may be larger than the
internal energy.  Nevertheless, the internal kinetic energy
represents a minimum value for the energy of the event that gave
rise to the the mushroom cloud.

The velocity gradient within the stem of the mushroom cloud could
be produced by accelerated streaming motion along the stem if the
stem is tilted at at constant angle along its length with respect
to the line of sight.  The same effect could be produced without
acceleration if the stem is the result of an ejection of material
with a velocity range of order 10 \kms \ (higher velocity gas
would appear furthest from the base). Also, a gradient can be
produced by a constant velocity flow with a monotonically
changing inclination angle along the length of the stem.  The
data at hand may not allow us to resolve this ambiguity.

Preliminary analysis of HI absorption from background continuum
sources through the cap indicate hydrogen spin temperatures of 50
to 100 K.  The cap contains dust emission which is visible in all
4 IRAS passbands, becoming more apparent with increasing
wavelength.  There is no obvious diffuse soft X-ray emission in
the energy range 0.1-2.0 keV, associated with any component of
the cloud (S. Snowden, private communication). X-ray emission
from hot gas that may be interior to the mushroom would be
attenuated via absorption by the neutral gas along the line of
sight.  Using the interstellar photoelectric absorption
cross-section of Morrison and McCammon (1983) and a neutral
hydrogen column density of $6\times10^{20}$ cm$^{-2}$, X-ray
emission from gas with a temperature less than $10^7$ K will be
absorbed. Thus the lack of observed X-rays appears to rule out the
existence of interior gas hotter than this.

\section{Discussion of Possible Scenarios}
\label{cfmodel}

The mushroom cloud shape and mass distribution of GW123.4-1.5 pose a
challenge to conventional superbubble scenarios. In these models
(e.g., see \nocite{macnor89} Mac Low, McCray, \& Norman 1989;
\nocite{GTTrozbod90} Tenorio-Tagle, Rozyczka, \& Bodenheimer 1990) the
lower part (the stem) retains the bulk of the mass, even though the
upper part of the bubble (the cap) may expand to a large radius.  This
is not the case with GW123.4-1.5, where we estimate that the cap
contains about four times the mass in the stem.  Additionally the
greatest cap to stem width ratio in the superbubble models is about
3:1 \nocite{GTTrozbod90} (Tenorio-Tagle, Rozyczka, \& Bodenheimer
1990) while the mushroom cloud's ratio is 6:1.  Furthermore, the
radius of the {\em model} stem is typically equal to $2 H$, where $H$
is the exponential scale length of the local Gaussian density
distribution; the stem is the cavity created by a blowout from a
stratified atmosphere into a uniform low-density halo (at $\sim$500
pc).  If the local scale length were equal to the global average of
the HI disk $H \simeq 135$ pc (\nocite{1990ARA&A..28..215DiLoc} Dickey
\& Lockman~1990), then the width of the mushroom's stem would be 500
pc rather than the $\sim$ 40 pc observed.

The above discrepancies lead us to consider alternate scenarios
in which a stem plus cap morphology can be realized.  For
example, an H~I jet could be ejected from the disk and a wide
lobe could be created where it has stalled, possibly falling back
to the disk in a fountain-like manner.  One possible origin for
such an event is the passage of a High Velocity Cloud through the
galactic disk and the subsequent emergence of gas on the other
side (\nocite{GTTfrabodroz87}Tenorio-Tagle et~al. 1987).
However, here we consider the rise of buoyant gas.  A rising
fireball after a terrestrial nuclear explosion creates a
structure strikingly similar to that of GW123.4-1.5.  In the
interstellar context, mushroom-shaped clouds can arise from the
interaction between different components in a multiphase medium
(e.g. \nocite{RosBreg95} Rosen \& Bregman 1995, \nocite{avil99}
Avillez 1999), including a cloud-cloud collision (Miniati et al.
1997).  For our model, we focus on the rise of buoyant hot gas
resulting from a single supernova event.  Jones (1973) has
investigated the early evolution of a supernova remnant and found
signs of buoyant rise. Here, we follow to late times the
evolution of a remnant which does not have enough energy to blow
out of the disk atmosphere. The initially pressure driven hot
bubble stalls at radius smaller than the scale height of the
medium\footnote{The outer shock front continues to propagate up
  through the stratified atmosphere, but has little dynamic
  importance since the inner hot bubble has stalled and is no
  longer compressing matter into a thin shell behind the shock
  front.}.  Buoyancy forces lift the low density bubble out of
the galactic plane and through the stratified atmosphere.  In
analogy with the nuclear fireball, the quasi-vacuum produced
under the rising cap may pull in surrounding material, entraining
it to form the stem. In this case, the stem width is not
determined by the scale height of the medium and can be quite
narrow.  However the existence of cold neutral material in the
cap must still be explained since the rise of a buoyant plume in
pressure equilibrium does not physically move much material from
the galactic disk to high latitudes.

To examine the plausibility of the buoyant bubble scenario we
have begun modeling a single 10$^{51}$ ergs explosion at 60 pc
above the midplane using a modified Zeus 2-d code (e.g.
\nocite{StoNor92} Stone \& Norman~1992).  Our simulations include
the effects of radiative cooling, heat conduction, and the
vertical gravitational field.  We also use artificially low
quiescent gas density to reduce the influence of ``numerical
diffusion", a computational artifact resulting from steep
gradients.  For example, we use a scale height 60 pc\footnote{The
  stalling radius is fixed for a given energy input and ambient
  density in the plane. So for any given source the buoyancy
  condition, that the stalling radius is less than $H$, is more
  likely to exist if the z-distribution of the ambient gas is $H
  \simeq 135$ pc .}  and a midplane number density of $\rm n_o =
1 \ atom \ cm^{-3}$ to create an ambient medium distribution
which allows bubbles to form and evolve.
 
Our preliminary low-resolution models are intentionally generic
and detailed comparisons with GW123.4-1.5 will be left to a later
paper.  In both Gaussian and exponential atmospheres, a bubble
forms interior to the blastwave and, rather than elongating
vertically as in a conventional superbubble, the bubble rises
buoyantly. Since the rise is supersonic relative to the cooled
gas above (interior to the blastwave), the bubble accumulates
cold ambient gas in a snow-plow mode along its shock front.  Gas
also flows upwards in a column following the bubble, contributing
to its evolution into a mushroom-shaped cloud.  By 8 Myr the
interior bubble has the kinetic temperature distribution shown in
Fig.~\ref{figmodel}a.  The models show a cold gas stem which is
narrower than the primary bubble (stem) in conventional
superbubble scenarios in which the stem width is dictated by the
gas scale height.  Our modeled stem is expected to become
narrower and more obvious than shown here when a higher ambient
density is used and cooling becomes more efficient. The coolest
temperatures trace a curlyqued cap structure like that observed
in GW 123.5-1.5. These structures are also evident in the column
density map, plotted for gas $<$7500 K, in Fig.~\ref{figmodel}b.
Similar to some models of supernova evolution
(\nocite{SlaCox92}Slavin \& Cox~1992) the hottest gas in our
simulation decreases to fill a small volume which eventually
collapses.  Residing just inside the skin of the lobes are
remnants of a warm gas envelope, formed by heat conduction, which
had been pushed away from the bubble by the upward flow of cold,
dense gas from the stem. Their temperature of
a few $\times 10^4$ K is consistent with the lack of soft x-ray
emission. The cool, H~I by-product of the buoyant bubble should
be observable for substantially longer than the cooling time
($\sim 2 \times 10^5$~yr) of this hot gas.

We have carried out a preliminary search in other wavebands for
evidence of a possible energy source, but no obvious candidates
can be identified. For example, there are no IRAS sources near
the base with CO emission in the FCRAO database (between the
midplane and - 3$^o$ galactic latitude) within the FWHM velocity
range of the stem.  The projected position of the H~II emitting
reflection nebula Sharpless~185 lies near the base of the stem.
However, it is associated with the Be X-ray emitting star
$\gamma$ Cass at a distance of only about two hundred pc (see
\nocite{bloudew97}Blouin et~al.~1997).

\newpage 

\figcaption[EngTayIrw.fig1.ps]{1a) Brightness temperature
  integrated over the range -31 \kms through -43 \kms . 
  The greyscale is an arbitrary stretch between 
  35 K \kms \ and 1460 K \kms .
  1b) The position-velocity plot for the longitude slice, along 
  the stem, marked in 1a.  1c) The position-velocity plot for 
  the latitude slice across the lobes. 1d) The position-velocity 
  plot for the latitude slice across the mid-region of the cap.
\label{obsmaps}}

\figcaption[EngTayIrw.fig2.ps]{2a) Temperature distribution for the
  8~Myr old remnant of a 10$^{51}$ erg explosion at 60~pc above
  the galactic plane in an exponential atmosphere with $H=60$~pc.
  In order to mimic the overall non-thermal pressure, the temperature
  of the undisturbed gas (region showing no velocity vectors) is
  isothermal with $T=10^4$~K.  Black lines with white edging mark
  supersonic shock front areas.  Both the initial blastwave and
  the shocks associated with the bubble rising supersonically are
  evident.  The cap is associated with the cold (black) region
  above the hot (white, curling) gas. The longest arrows correspond to
  flow velocities of $\sim 50$~km~s$^{-1}$. 2b) Our
  estimate of the column density ($10^{20}$~cm$^{-2}$), for
  gas with $T < 7500$~K, also increases
  in the cold region, delineating a cap, and has a narrower stem
  than appears in the distribution of slowly varying gas
  temperature displayed in 2a).
\label{figmodel}}

\newpage 
\tablenum{1}
\newcommand\cola {\null}
\newcommand\colb {&}
\newcommand\colc {&}
\newcommand\cold {&}
\newcommand\cole {&}
\newcommand\colg {&}
\newcommand\colh {&}
\newcommand\eol{\\}
\newcommand\extline{&&&&&&\eol}
\begin{deluxetable}{llccccc}
\tablewidth{0pc}
\tablecaption{Components of GW~123.4-1.5. \label{tablepar}}
\tablehead{
\multicolumn{2}{l}{Component}
\colc \multicolumn{3}{c}{Velocities}
\colg {Mass\tablenotemark{a}}
\colh {Energy}
\eol
\cola {} \colb {}
\colc \multicolumn{3}{c}{(km s$^{-1}$)}
\colg {($\rm \times 10^3 \ \cal M_{\odot}$)}
\colh {($\rm \times 10^{50}$ Ergs)}
\eol
\extline
\cola {} \colb {}
\colc {Peak}  \cold{FWHM} 
\cole  {Range} \colg{ }
\colh {Kinetic\tablenotemark{b}}
\eol 
}
\startdata
 \cola {STEM} \colb  
   \colc -43 $\pm$ 2\cold 9 $\pm$ 1\cole -49.5 to -36.9
      \colg {\bf 35} 
   \colh{\bf 0.1}
\eol
 \cola {CAP}\colb  
   \colc -35.3 $\pm$ 0.6\cold 9 $\pm$ 1\cole -44.8 to  -25.8
     \colg {\bf 120 }
\colh{\bf 1.1 }
\eol
 \cola { }\colb {lobes}
   \colc -39.7 $\pm$ 0.6\cold 8 $\pm$ 1\cole -44.8 to -30.3
      \colg 40 \colh
\eol
 \cola { }\colb {mid-region}
   \colc -35.7  $\pm$ 0.6 \cold 9 $\pm$ 1\cole -44.8 to -30.3
       \colg 55 \colh
\eol
 \cola { }\colb {wisps}
   \colc -32.8 $\pm$ 0.6 \cold 8 $\pm$ 1\cole -36.1 to -25.8 
      \colg 25 \colh
\eol
 \cola {WHOLE}\colb  
   \colc -37 $\pm$ 1 \cold 12 $\pm$ 3\cole -49.5 to -25.8 
       \colg {\bf 155 }
\colh{\bf 2 }
\eol
\tablenotetext{a}{Calculated for a distance of 3.8 kpc. The
inaccuracy associated with the 
distance results in 
an uncertainty of about 50\% for the mass. 
features were excluded from these
estimates:  the diagonal features (near galactic coordinates
125\adeg \ 23\amin \ 29.7\asec 
\ -5\adeg \ 13\amin \ 30\asec),
the H~I ring (surrounding 123\adeg \ 30\amin \ 50.1\asec 
\ -5\adeg \ 31\amin \ 33\asec), and Sharpless 184 (near
\ 123\adeg \ 10\amin \ 12\asec \ -6\adeg \ 17\amin \ 8\asec).}
\tablenotetext{b} {Calculated using half the velocity range. The
value for the whole structure includes internal and relative 
motions.}
\enddata
\end{deluxetable}

\end{document}